\documentclass[twocolumn,showpacs,preprintnumbers,amsmath,amssymb]{revtex4}
\usepackage{amssymb}
\usepackage{txfonts}

\usepackage{graphicx}
\usepackage{dcolumn}
\usepackage{bm}
\usepackage{sidecap}

\emergencystretch=\maxdimen
\begin{document}

\title{Non-Markovian master equation in strong-coupling regime}

\author{Yang Li$^{1}$}
\author{Bin Luo$^{1}$}
\author{Jian-Wei Zhang$^{2}$}
\author{Hong Guo$^{1}$}\thanks{Correspondence author.
Phone: +86-10-6275-7035,
Fax: +86-10-6275-3208,
E-mail: hongguo@pku.edu.cn.}

\affiliation{$^{1}$ CREAM Group, State Key Laboratory of Advanced Optical Communication Systems and Networks
(Peking University),
School of Electronics Engineering and Computer Science,
and Center for Computational Science and Engineering (CCSE),
Peking University, Beijing 100871, China\\
$^{2}$ School of Physics, Peking University, Beijing 100871, China}%

\date{\today}

\begin{abstract}
The time-convolutionless (TCL) non-Markovian master equation was generally thought to break down at finite time due to its singularity
and fail to produce the asymptotic behavior in strong coupling regime. However, in this paper, we show that the singularity is not an obstacle for validity of the TCL master equation. Further, we propose a multiscale perturbative method valid for solving the TCL master equation in strong coupling regime, though the ordinary perturbative method invalidates therein.
\end{abstract}

.

\maketitle

\section{Introduction}
The study of non-Markovian quantum open system attracts increasing attention nowadays. There are two reasons for the interest.
On the one hand, the popular Markovian approximation which neglects the memory effects of the environment is not sufficient
for the recent progress in many fields,
such as quantum information processing \cite{QI1}, quantum optics \cite{QO1,QO2,QO3}, condensed matter physics \cite{cond1,cond2},
chemical physics \cite{chem}, and even life science \cite{bio}. On the other hand, there are still many open
questions for the theory of non-Markovian quantum open system.

Several methods are proposed to study the non-Markovian open quantum systems \cite{QO1,NM1,NM2,NM3,NM4}, among which
the non-Markovian master equation is quite promising. Projection operator techniques provide systematic framework to derive master equations.
Different projection operator techniques give different kinds of non-Markovian master equations. Two kinds of non-Markovian master equations,
the Nakajima-Zwanzig master equation \cite{NM1} and the time-convolutionless (TCL) master equation \cite{NM2}, are widely used.
Since the Nakajima-Zwanzig master equation is an integro-differential equation, the TCL master equaiton which is a time-local first order differential
equaiton is much easier for numerical solution. Besides the methods by extending the Hilbert space \cite{extend1,extend2}, a new numerical method
called non-Markovian quantum jump \cite{NMQJ0} and its modified scheme \cite{NMQJ} were proposed recently.

However, in strong-coupling regime,
it was generally thought that there are two severe problems for the application of the TCL master equation.
One is that the TCL master equation breaks down at finite time in strong-coupling regime due to the singularity,
thus fails to produce the asymptotic behavior \cite{QO1,GME}.
Another is that the ordinary perturbative method fails to produce the correct behavior \cite{QO1}.

In this Letter, we study the dynamics of a two-level system interacting
with a structured environment in strong-coupling regime.
Our result shows that
the singularity at finite time is not an obstacle for the TCL master equaiton to produce the correct asymptotic behavior.
Moreover, we introduce a multiscale perturbative method which produces the correct behavior, though the ordinary perturbative method fails, in strong-coupling regime.

\section{Theoretical framework}
Under rotating wave approximation,
the total Hamiltonian of the two-level system with the bosonic reservoir
in zero-temperature
is given by
$
H = H_S  + H_E  + H_I  = H_0  + H_I\,,
$
with
$H_S  = \omega _0 \sigma _ +  \sigma _ -\,,  H_E  = \sum_k {\omega _k a_k^\dag  a_k }\,,H_I  = \sum_k {(g_k \sigma _ +  a_k  + g_k^* a_k^\dag  \sigma _ -  )}\,,$
($\hbar  = 1$).
Here $\sigma _ \pm$ and $\omega_0$ are the inversion operators
and
transition frequency of the two-level system, respectively,
$a_k^\dag$, $a_k$
the creation and annihilation operators of the field modes of the reservoir
with frequency $\omega_k$, and $g_k$,
$g_k^*$
the coupling strength between the two-level system and the $k$th field mode of the reservoir. Since $[H, N]=0$,
where $N=\sigma_+ \sigma_- +\sum_k {a_k^\dag a_k}$,
for an initial state of the form
$|\ {\psi (0)} \rangle  = (c_{g0} |\ g \rangle  + c_{e0} |\ e \rangle )|\ 0 \rangle _E$,
the time evolution of the total system
is confined to the subspace spanned by the bases
$\{ |\ g \rangle |\ 0 \rangle _E ,|\ e \rangle |\ 0 \rangle _E ,|\ g \rangle |\ {1_k } \rangle _E \}$
as
\begin{eqnarray}
|\ {\psi (t)} \rangle  = c_{g0} |\ g \rangle |\ 0 \rangle _E  + c_e (t)|\ e \rangle |\ 0 \rangle _E  + \sum_k {c_k (t)|\ g \rangle |\ {1_k } \rangle _E }\,,
\label{wf}
\end{eqnarray}
\noindent where $ |\ {1_k }  \rangle _E$ is the state of the reservoir with only one exciton in the $k$th mode.

According to the Schr$\rm{\ddot{o}}$dinger equation
in the interaction picture with $H_0=H_S+H_E$, one can obtain an integro-differential equation for the amplitude $c_e (t)$ as
\begin{eqnarray}
\label{eqce}
\dot c_e (t) =  - \int_0^t\! {d\tau f(t - \tau )\ c_e (\tau)}\,,
\end{eqnarray}
where the correlation function $f(t-\tau)$ takes the form
\begin{eqnarray}
f(t - \tau ) = \sum_k { |\ {g_k }  |^2 e^{i(\omega _0  - \omega _k )(t - \tau )} }  = \int_0^\infty \! {J(\omega _k )e^{i(\omega _0  - \omega _k )(t - \tau )} } d\omega_k ,
\nonumber
\end{eqnarray}
where $J(\omega_k)$ is the spectral density function of the reservoir.

\noindent Thus, the TCL master equation takes the following form \cite{QO1,GME,GME2}
\begin{eqnarray}
\label{TCL}
\frac{d}{{dt}}\rho_s (t) &=&  - \frac{i}{2}S(t)[\sigma _ +  \sigma _ -  ,\rho_s (t)]  \nonumber\\
&& + \gamma (t)[\sigma _ -  \rho_s(t) \sigma _ +   - \frac{1}{2}\{ \sigma _ +  \sigma _ -  ,\rho_s(t) \} ]\,,
\end{eqnarray}
with Lamb shift and decay rate given by
\begin{eqnarray}
\label{SG}
S(t) =  - 2{\mathop{\rm Im}\nolimits}  \left[ {\frac{{\dot c_e (t)}}{{c_e (t)}}}  \right],{\kern 1pt} {\kern 1pt} {\kern 1pt} \gamma (t) =  - 2{\mathop{\rm Re}\nolimits}  \left[ {\frac{{\dot c_e (t)}}{{c_e (t)}}}  \right]\,.
\end{eqnarray}

\begin{figure}
\centering
\includegraphics[height=0.35\textheight,width=0.5\textwidth]{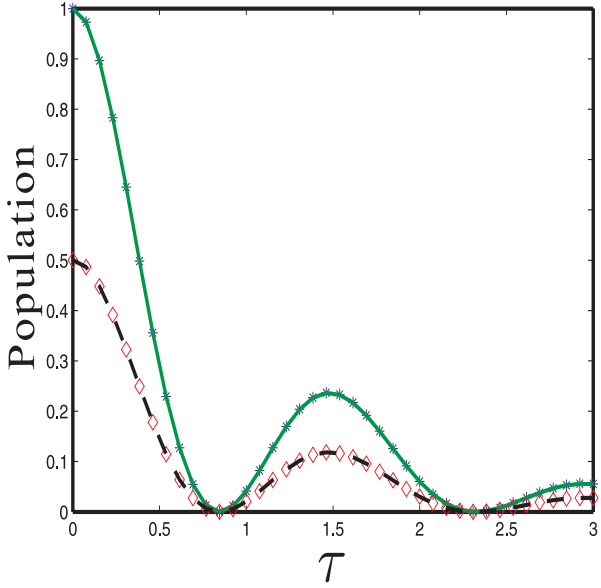}
\caption{\small (Color online) Time evolution of the population in the upper level for different initial states
(a) $ |\ {\phi_1 }  \rangle  =  |\ e  \rangle$
and
(b) $ |\ {\phi_2 }  \rangle  = ( |\ e  \rangle  +  |\ g  \rangle )/\sqrt 2\,,$  ($\tau =\lambda t$),
both with
$\lambda=1$, $\gamma_0=10$,
for the cases of
(i) the numerical solution of TCL master equation for initial state
$ |\ {\phi_1 } \rangle  =  |\ e  \rangle$ (asterisk),
(ii) the exact analytical solution for initial state $ |\ {\phi_1 }  \rangle  =  |\ e  \rangle$ (solid curve),
(iii) the numerical solution of TCL master equation for initial state $ |\ {\phi _2 }  \rangle  = ( |\ e  \rangle  +  |\ g  \rangle )/\sqrt 2$ (diamond),
and
(iv) the exact analytical solution for initial state $ |\ {\phi_2 }  \rangle  = ( |\ e  \rangle  + |\ g  \rangle )/\sqrt 2$ (dashed curve). } \label{fig1}
\end{figure}

\subsection{Validity of TCL master equation}
One problem for the TCL master equation is that it was thought to break down at finite time in strong-coupling regime due to the singularity.
For example, the problem occurs in the case that the spectral density of the reservoir takes the Lorentzian form
$J(\omega _k ) = \gamma _0 \lambda ^2 /2\pi [(\omega _k  - \omega _c )^2  + \lambda ^2 ]$
and
the two-level system interacts with the central frequency of the reservoir resonantly,
$\omega_0 = \omega_c$ \cite{QO1,GME}.
In the following, we restudy this problem for this model.
If $\omega_c \gg \lambda$, such as in an optical cavity,
$\omega_c$ can be extended to infinity.
Then the correlation function $f(t-\tau)$ is given by
\begin{eqnarray}
\label{ft}
f(t - \tau ) = \frac{{\gamma _0 \lambda }}{2}e^{ - \lambda  | {t - \tau }  |}\,.
\end{eqnarray}
Substituting Eq.~(\ref{ft}) into (\ref{eqce}) and using Laplace approach,
one obtains \cite{QO1}
\begin{eqnarray}
\label{ce}
c_e (t) = c_{e0} e^{ - \lambda t/2}  \left[ {\cos  \left({\frac{{\Gamma}}{2} t }  \right)+ \frac{\lambda }{\Gamma}\sin  \left({\frac{{\Gamma}}{2} t }  \right)}  \right]\,,
\end{eqnarray}
where
$\Gamma = \sqrt {2\gamma _0 \lambda  - \lambda ^2 }$.
For the strong-coupling case
($\gamma_0 > \lambda /2$),
$\Gamma > 0$, and $c_e (t)$
is
an oscillating function with discrete zeros
at
$t_n = 2  [ {(n + 1)\pi  - \arctan (\Gamma /\lambda) }  ]/ \Gamma$, ($n = 0, 1, 2, \cdots$).
Substituting Eq. (\ref{ce}) into (\ref{SG}),
one obtains the exact expressions for $S(t)$ and $\gamma (t)$ \cite{GME}
\begin{eqnarray}
\label{exact}
S(t) = 0, \ \ \gamma (t) = \frac{2\gamma _0\tan(\Gamma t/2)}{1 + \lambda/\Gamma\tan(\Gamma t/2)}.
\end{eqnarray}
Therefore,
we see that
$\gamma (t)$ diverges at these points $t_n$.

Previously, it was generally thought that the singularity is an obstacle for validity of the TCL master equation \cite{QO1,GME}.
On one hand, for TCL master equation, since ``the evolution of the reduced density matrix
only depends on the actual value of $\rho_s (t)$ and on the TCL generator'' \cite{QO1} and the density matrices coincide
at $t=t_0$ for different initial states, the evolution of the density matrices after $t_0$
{\it should} be the same for different initial states.
On the other hand, the exact analytical solution \cite{QO1} for the problem shows that
the density matrices for $t> t_0$ differs for different initial states.
That means the solution of TCL master equation {\it does not agree with} the exact analytical solution for $t> t_0$.
So, it was thought that,
``...a time-convolutionless form of the equation of motion which
 is local
 in time ceases to exist for $t>t_0$...'' \cite{QO1}
or ``...the time-convolutionless generator breaks down
 at finite time in the strong coupling regime,
 thus failing to reproduce the asymptotic behavior...'' \cite{GME}.

In our opinion, the singularity is {\it not} an obstacle for validity of the TCL master equation.
By using the method in \cite{NMQJ},
we solve the TCL master equation Eq. (\ref{TCL}) numerically. In the simulation, the decay rate (Eq. (\ref{exact})) with
singularity is used. From Fig. \ref{fig1}, we find that the numerical solution of TCL master equation agrees with the exact analytical solution very well for $t>t_0$.
That means, even though the TCL master equation has
a
singular point at $t=t_0$ in strong-coupling regime,
it still reproduces the correct dynamics when $t>t_0$.
Actually, since $t=t_0$ is a singulary point of the TCL master equation,
the dynamics around $t=t_0$ cannot be explained by the theory of the first-order ordinary differential equation at an ordinary point \cite{ODE}.

\subsection{Multiscale perturbative expansion}
Another
crucial
problem for TCL master equation is that
 the ordinary perturbative expansion fails in strong-coupling regime \cite{QO1}.
The reason is that
 the ordinary perturbative expansion corresponds basically to a Taylor expansion of $\gamma (t)$
  in powers of $\gamma_0$.
In fact, this method treats the dynamics only in one time scale.
For the model considered above in strong-coupling regime,
there are two time scales, which correspond to
 the decaying
 and
 the oscillating behaviors, respectively.
The ordinary perturbative expansion only considers
 the time scale of the decaying behavior,
 so the oscillating behavior disappears in the perturbative solution (see Fig. 2).

\begin{figure}
\centering
\includegraphics[height=0.35\textheight,width=0.5\textwidth]{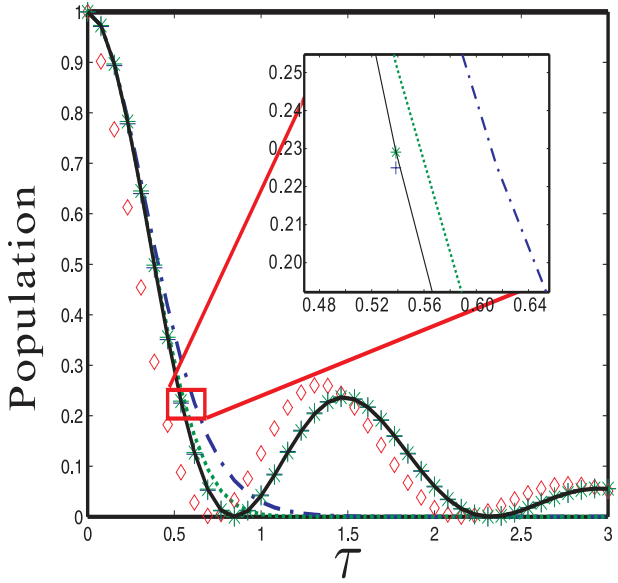}
\caption{\small (Color online) Time evolution of the population in the upper level for initial state $ |\ e  \rangle$
by multiscale perturbative method and ordinary perturbative method ($\tau =\lambda t$),
 with $\lambda=1$, $\gamma_0=10$,
 for the cases of
(i) the exact solution (solid curve),
(ii) the numerical solution of exact TCL master equation (cross),
(iii) the first order multiscale method (diamond),
(iv) the second order multicale method (asterisk),
(v) the second order ordinary perturbative method (dash-dotted curve),
(vi) the fourth order ordinary perturbative method (dotted curve). } \label{fig2}
\end{figure}

Since the failure of the ordinary perturbative expansion originates from
 ignoring multiscales of the dynamics,
we introduce a multiscale perturbative expansion \cite{Multi} to treat the strong-coupling case.

According to Eq.~(\ref{SG}), by giving a multiscale perturbative expansion
of $c_e (t)$, one can get the multiscale perturbative expansions of $\gamma (t)$ and $S(t)$.
From
Eqs.~(\ref{eqce}) and (\ref{ft}), one obtains
\begin{eqnarray}
\label{eqce2}
\ddot c_e (t) + \lambda \dot c_e (t) + \frac{{\gamma _0 \lambda }}{2}c_e (t) = 0\,.
\end{eqnarray}
By introducing dimensionless parameters $T=\gamma_0 t$
 and
 $\varepsilon {\rm{ = }}\lambda {\rm{/}}\gamma _0$,
 where
 $\varepsilon \ll 1$ for strong-coupling regime,
Eq.~(\ref{eqce2})
reads
\begin{eqnarray}
\label{eqce3}
\frac{{d^2 }}{{dT^2 }}c_e  + \varepsilon \frac{d}{{dT}}c_e  + \frac{\varepsilon }{2}c_e  = 0\,,
\end{eqnarray}
with initial conditions $c_e(0)=c_{e0}$ and $\dot c_{e}(0)=0$.
There are two kinds of behaviors of the dynamics,
corresponding to
 two different time scales.
The time scale of the decaying behavior relates to
$\lambda t= \varepsilon T$,
and that of
the oscillating behavior relates to a complicated function of $\varepsilon$.
Therefore, two different time scales,
 $t_1 = (\varepsilon ^{1/2} a_1 + \varepsilon ^{3/2} a_2+ ...)T$
 and
 $t_2 = \varepsilon T$, are introduced,
  where $a_i$'s are unknown parameters to be determined.
Expanding $c_e (t_1, t_2)$ in powers of $\varepsilon$
\begin{eqnarray}
\label{cem}
c_e (t_1, t_2)= c_e ^{(0)}(t_1, t_2) + \varepsilon ^{1/2} c_e ^{(1)}(t_1, t_2) + \varepsilon c_e ^{(2)}(t_1, t_2) +...\,,
\end{eqnarray}
and substituting Eq.~(\ref{cem}) into (\ref{eqce3})
and the initial conditions,
one obtains the equations for $c_e ^{(i)}$ as follows.

To the order of $\varepsilon$, one gets the equations for $c_e ^{(0)}$ as
\begin{eqnarray}
&&a_1^2 \frac{{\partial ^2 }}{{\partial t_1^2 }}c_e^{(0)}  + \frac{1}{2}c_e^{(0)}  = 0\,,
\nonumber \\
&&c_e^{(0)} (0,0) = c_{e0} , \ \ \frac{\partial }{{\partial t_1 }}c_e^{(0)} (0,0) = 0\,.
\nonumber
\end{eqnarray}

To the order of $\varepsilon ^{3/2}$, one gets the equations for $c_e ^{(1)}$ as
\begin{eqnarray}
&&a_1^2 \frac{{\partial ^2 }}{{\partial t_1^2 }}c_e^{(1)}  + \frac{1}{2}c_e^{(1)}  =  - 2a_1 \frac{{\partial ^2 }}{{\partial t_1 \partial t_2 }}c_e^{(0)}  - a_1 \frac{\partial }{{\partial t_1 }}c_e^{(0)}\,,
\nonumber \\
&&c_e^{(1)} (0,0) = 0\,,
\nonumber \\
&&a_1 \frac{\partial }{{\partial t_1 }}c_e^{(1)} (0,0) + \frac{\partial }{{\partial t_2 }}c_e^{(0)} (0,0) = 0\,.
\nonumber
\end{eqnarray}

To the order of $\varepsilon ^{2}$, one gets the equations for $c_e ^{(2)}$ as
\begin{eqnarray}
&&a_1^2 \frac{{\partial ^2 }}{{\partial t_1^2 }}c_e^{(2)}  + \frac{1}{2}c_e^{(2)}
 = - 2a_1 \frac{{\partial ^2 }}{{\partial t_1 \partial t_2 }}c_e^{(1)}
 - 2a_1 a_2 \frac{{\partial ^2 }}{{\partial t_1^2 }}c_e^{(0)}
\nonumber\\
&& \qquad   - \frac{{\partial ^2 }}{{\partial t_2^2 }}c_e^{(0)}  - a_1 \frac{\partial }{{\partial t_1 }}c_e^{(1)}  - \frac{\partial }{{\partial t_2 }}c_e^{(0)}\,,
\nonumber\\
&&c_e^{(2)} (0,0) = 0\,,
\nonumber\\
&&a_1 \frac{\partial }{{\partial t_1 }}c_e^{(2)} (0,0) + a_2 \frac{\partial }{{\partial t_1 }}c_e^{(0)} (0,0) + \frac{\partial }{{\partial t_2 }}c_e^{(1)} (0,0) = 0\,.
\nonumber
\end{eqnarray}

By a routine multiscale analysis of the above equations \cite{Multi}, one obtains the solutions of $c_e ^{(i)}$.
By substituting the perturbative solution of $c_e$ into Eq. (\ref{SG}), we can get the corresponding Lamb shifts and decay rates.
The first order solution is
\[
S(t) = 0, \ \
\gamma (t) = \varepsilon  + \sqrt {2\varepsilon } \tan (\sqrt {2\varepsilon } T/2),
\]
and the second order solution is
\begin{widetext}
\[
S(t) = 0, \ \ \ \ \ \ \ \ \ \ \ \
\gamma (t) = \frac{{\sqrt \varepsilon  \{ \varepsilon ^{3/2} \cos [\sqrt {2\varepsilon } (1 - \varepsilon /4)T/2] + \sqrt 2 (4 + \varepsilon )\sin [\sqrt {2\varepsilon } (1 - \varepsilon /4)T/2]\} }}{{4\cos [\sqrt {2\varepsilon } (1 - \varepsilon /4)T/2] + 2\sqrt {2\varepsilon } \sin [\sqrt {2\varepsilon } (1 - \varepsilon /4)T/2]}}.
\]
\end{widetext}

By solving the corresponding master equations using the method in \cite{NMQJ},
we study the dynamics of the population in the upper level.
From Fig. \ref{fig2}, we can see that,
unlike the solutions of ordinary perturbative method where the oscillating behavior is missing,
the decaying and oscillating behaviors are both included by multiscale
perturbative solutions. The perturbative solution up to the second order fits very well with the exact solution.

For the exact solution, Eq.~(\ref{exact}),
 the first singular time of $\gamma(t)$ is $t_0  = 2\arccos ( - \sqrt {\varepsilon /2} )/[\gamma_0 \sqrt {(2 - \varepsilon )\varepsilon }]$.
For the first and second order approximations of $t_0$ by the multiscale perturbative expansion,
the first singular time of $\gamma(t)$ are $\pi /[\gamma_0 \sqrt {2\varepsilon }]$ and $4\sqrt 2 \arccos [ - \sqrt {\varepsilon /(2 + \varepsilon )} ]/[\gamma_0 (4 - \varepsilon )\sqrt \varepsilon]$, respectively.
Detailed analysis shows that relative errors of the first and second order approximations are in the orders of $\varepsilon ^{1/2}$ and $\varepsilon ^{3/2}$, respectively.

\section{Conclusion}
To summarize, we study the time-convolutionless
non-Markovian master equation in strong-coupling regime.
For the environment with Lorentzian spectral density,
we find that the singularity at finite time does not
influence the master equation to produce the correct asymptotic behavior of the open system.
We also propose a multiscale perturbative method, which fits well with the exact solution, though the ordinary perturbative method fails, in strong-coupling regime.

\section*{ACKNOWLEDGMENTS}
We thank H.M.\ Wiseman, W.T.\ Strunz, An Jun-Hong and C.J.\ Wu for fruitful discussions. This work is supported by the Key Project of the National Natural Science Foundation of China (Grant No. 60837004) and National Hi-Tech Research and Development Program (863 Program).


\end{document}